\documentclass[%
 reprint,
superscriptaddress,
 amsmath,amssymb,
prb,
]{revtex4-2}

\usepackage{graphicx}
\usepackage{dcolumn}
\usepackage{bm}
\usepackage{xcolor}

\newcommand{\uspace}{\(\mskip3mu\)} 

\begin{document}

\title{THz ultra-strong light-matter coupling up to 200K with continuously-graded parabolic quantum wells}

\author{Paul Goulain}
\affiliation{Centre de Nanosciences et de Nanotechnologies, CNRS UMR 9001, University of Paris-Saclay, 91120 Palaiseau, France}

\author{Chris Deimert}
\affiliation{Department of Electrical and Computer Engineering, University of Waterloo, 200 University Ave W, Waterloo, ON N2L 3G1, Canada}

\author{Mathieu Jeannin}
\affiliation{Centre de Nanosciences et de Nanotechnologies, CNRS UMR 9001, University of Paris-Saclay, 91120 Palaiseau, France}

\author{Stefano Pirotta}
\affiliation{Centre de Nanosciences et de Nanotechnologies, CNRS UMR 9001, University of Paris-Saclay, 91120 Palaiseau, France}

\author{Wojciech Julian Pasek}
\affiliation{Faculdade de Ciências Aplicadas, Universidade Estadual de Campinas, Limeira, SP, 13484-350, Brazil}

\author{Zbigniew Wasilewski}
\affiliation{Institute for Quantum Computing, University of Waterloo, 200 University Ave W, Waterloo, ON N2L 3G1, Canada}
\affiliation{Waterloo Institute for Nanotechnology, University of Waterloo, 200 University Ave W, Waterloo, ON N2L 3G1, Canada}
\affiliation{Department of Physics and Astronomy, University of Waterloo, 200 University Ave W, Waterloo, ON N2L 3G1, Canada}

\author{Raffaele Colombelli}
\email{raffaele.colombelli@universite-paris-saclay.fr}
\affiliation{Centre de Nanosciences et de Nanotechnologies, CNRS UMR 9001, University of Paris-Saclay, 91120 Palaiseau, France}

\author{Jean-Michel Manceau}
\email{jean-michel.manceau@universite-paris-saclay.fr}
\affiliation{Centre de Nanosciences et de Nanotechnologies, CNRS UMR 9001, University of Paris-Saclay, 91120 Palaiseau, France}


\begin{abstract}

  Continuously graded parabolic quantum wells with excellent optical performances are used to overcome the low-frequency and thermal limitations of square quantum wells at terahertz frequencies.
  The formation of microcavity intersubband polaritons at frequencies as low as 1.8\uspace THz is demonstrated, with a sustained ultra-strong coupling regime up to a temperature of 200\uspace K. Thanks to the excellent intersubband transition linewidth, polaritons present quality factors up to 17.
  It is additionally shown that the ultra-strong coupling regime is preserved when the active region is embedded in sub-wavelength resonators, with an estimated relative strength \(\eta = \Omega_R / \omega_0 = 0.12\). 
  This represents an important milestone for future studies of quantum vacuum radiation because such resonators can be optically modulated at ultrafast rates, possibly leading to the generation of non-classical light via the dynamic Casimir effect.
  Finally, with an effective volume of \(2.10^{-6} \lambda_0^3\), it is estimated that fewer than 3000 electrons per resonator are ultra-strongly coupled to the quantized electromagnetic mode, proving it is also a promising approach to explore few-electron polaritonic systems operating at relatively high temperatures. 

\end{abstract}

\maketitle

\section{Introduction}

Following the pioneering work by Haroche et al.\cite{kaluznyObservationSelfInducedRabi1983}, engineering of the light-matter interaction has been one of the most active fields of modern physics.
Remarkably, when the losses in the system are lower than the coupling strength, a new regime is attained - the strong coupling regime (SCR) - where light and matter exchange energy coherently and periodically.
In the frequency domain, this leads to a radical change of the system's spectral response.
From the first observation with Rydberg atoms, the SCR has been demonstrated in a plethora of systems spanning excitons, organic molecules, electronic transitions, superconducting qubits and many others.\cite{basovPolaritonPanorama2021}
The strength of the light-matter interaction is often gauged by a dimensionless parameter \(\eta\) that is the ratio between the coupling constant \(\Omega_R\) (also called the vacuum Rabi frequency) over the resonant transition frequency \(\omega_0\).\cite{friskkockumUltrastrongCouplingLight2019}
Above a value \(\mathrm{\eta > 0.1}\), one enters the ultra-strong coupling (USC) regime where the diamagnetic terms of the interaction Hamiltonian start to play an important role, leading to a deviation from the linear approximation and the formation of a sizeable population of virtual photons in the ground state of the system.\cite{ciutiQuantumVacuumProperties2005}
The same foundational article by Ciuti et al.\cite{ciutiQuantumVacuumProperties2005} also proposed that an abrupt modulation of the system ground state leads to a release of such virtual population as real photons, an approach that could lead to the development of non-classical light emitters at long-wavelengths. 

The first candidate for this USC regime of interaction were intersubband (ISB) polaritons.\cite{ciutiQuantumVacuumProperties2005}
These hybrid states stem from the strong coupling between an ISB transition (more rigorously, an ISB plasmon\cite{andoElectronicPropertiesTwodimensional1982}) within the conduction band of doped semiconductor quantum wells and the quantized electro-magnetic mode of a microcavity.
They can operate down to the THz frequency range, while their Rabi frequency - proportional to the square root of the introduced dopant density - can reach values equivalent to the bare transition.
Soon after the first experimental demonstration\cite{diniMicrocavityPolaritonSplitting2003}, the USC regime with ISB polaritons was demonstrated in two different microcavity configurations.\cite{anapparaSignaturesUltrastrongLightmatter2009,todorovUltrastrongLightMatterCoupling2010}
Since then, USC was demonstrated with organic systems operating at visible wavelength and ambient temperature as thoroughly discussed in \cite{friskkockumUltrastrongCouplingLight2019}, and more recently with plasmonic resonances embedded in microcavity that could be in principle extended to the far infrared range.\cite{baranov_ultrastrong_2020} With doped heterostructures, the coupling strength was increased further using two different strategies that alter the character of the quantum well (QW) resonance.
In the first configuration, a massive increase of the doping density in QWs was found to lead to the formation of a multisubband plasmon\cite{alpeggianiSemiclassicalTheoryMultisubband2014}: a resonance essentially locked to the plasma frequency.
This enabled a coupling strength \(\mathrm{\eta = 0.47}\), once placed in a resonant metallic microcavity at mid-infrared wavelengths.\cite{askenaziMidinfraredUltrastrongLight2017}
In the second configuration, the application of a strong magnetic field to a 2D electron gas leads to the formation of Landau levels at sub-THz frequencies.
When coupled to sub-wavelength resonators, Landau polaritons form with a coupling strength that can reach values beyond unity.\cite{muravevObservationHybridPlasmonphoton2011,scalariUltrastrongCouplingCyclotron2012,bayerTerahertzLightMatter2017}
Several attempts have been made to obtain ultrafast modulation of the USC regime with ISB polaritons,\cite{gunterSubcycleSwitchonUltrastrong2009,zanottoUltrafastOpticalBleaching2012} with most of the activity aiming at measuring quantum vacuum radiation now conducted with Landau polaritons.\cite{halbhuberNonadiabaticStrippingCavity2020}
Although they offer coupling strengths around/beyond unity, they require elevated magnetic fields and cryogenic temperatures below 8\uspace K.

Two primary factors have prevented further increase of \(\eta\) with THz ISB polaritons - particularly in square QWs.
On one hand, any attempt to increase the coupling strength (\(\Omega_R\)) by increasing the doping results in a blueshift of the ISB transition, leaving the ratio \(\eta\) unchanged.
This is due to the Coulomb interactions that act as a self-energy correction on the optical transition.
This effect, commonly known as the depolarization shift,\cite{andoElectronicPropertiesTwodimensional1982,zaluznyInterSubbandOpticalAbsorption1982} is proportional to the square root of the sheet doping density.
On the other hand, any attempt to lower the transition frequency (\(\omega_0\)) while maintaining a high doping density will spread the population (\(\mathrm{n_{2D}}\)) over two subbands.
This leads to the activation of a second ISB transition and no further optimization of \(\eta\) is possible.
Careful simulations of a square QW with a doping density of 1.10\textsuperscript{11}\uspace cm\textsuperscript{-2}, using a commercial solver (NextNano), reveal that it is not possible to obtain a single and sharp ISB transition below 2.5\uspace THz.
Accessing the lower frequency range is, however, a prerequisite to demonstrate non-classical light generation via the dynamic Casimir effect.
The reason is that elucidating such subtle effects requires coherent field correlation measurements, which have so far been done using the electro-optic sampling technique,\cite{moskalenkoParaxialTheoryDirect2015,riekDirectSamplingElectricfield2015,benea-chelmusSubcycleMeasurementIntensity2016,benea-chelmusElectricFieldCorrelation2019} whose highest sensitivity bandwidth is in the 1-2\uspace THz range.

\section{Quantum wells with parabolic energy potential}

\subsection{Design and implementation}

A promising approach to circumvent this problem is to move from the standard square QW to parabolic energy potentials.
The key advantage is that the electron-electron interaction cancels out - analogous to the Kohn's theorem for cyclotron resonance\cite{kohnCyclotronResonanceHaasvan1961} - leaving the transition energy independent from distribution and density of the electrons.
Digitally graded parabolic energy potentials have already been used for the demonstration of strong light-matter coupling at THz frequencies, but they suffer low optical performances due to the interface roughness broadening that is inherent to the interdigitated epitaxial growth technique.\cite{geiserUltrastrongCouplingRegime2012,paulilloRoomTemperatureStrong2016}
Furthermore, these demonstrations were not designed to operate below the frequency of 2.5\uspace THz imposed by classical square QW structures.
A recent experimental tour de force has radically changed the perspective in that field.
In Ref.\cite{deimertMBEGrowthContinuouslygraded2019}, the parabolic energy potential is no longer mimicked by a series of narrow square QWs, but it is faithfully reproduced by a continuously graded alloy composition. This requires precise time-varying group III fluxes, which are challenging to produce in a conventional MBE setup, primarily because there is a thermal lag in the response of the group III cell. The approach developed by Deimert at al. in Ref.\cite{deimertMBEGrowthContinuouslygraded2019} uses a transfer function model to counteract the thermal dynamics, along with an additional corrective step to achieve high-precision composition gradients at typical growth rates.
This approach led to high performance THz ISB transitions with record linewidths up to 150\uspace K, and to stable operation up to room temperature.\cite{deimertRealizationHarmonicOscillator2020}
In the present letter, we build on that result and we experimentally show USC regime with ISB transitions below the inherent frequency limit of square QWs.

\begin{figure}
  \includegraphics{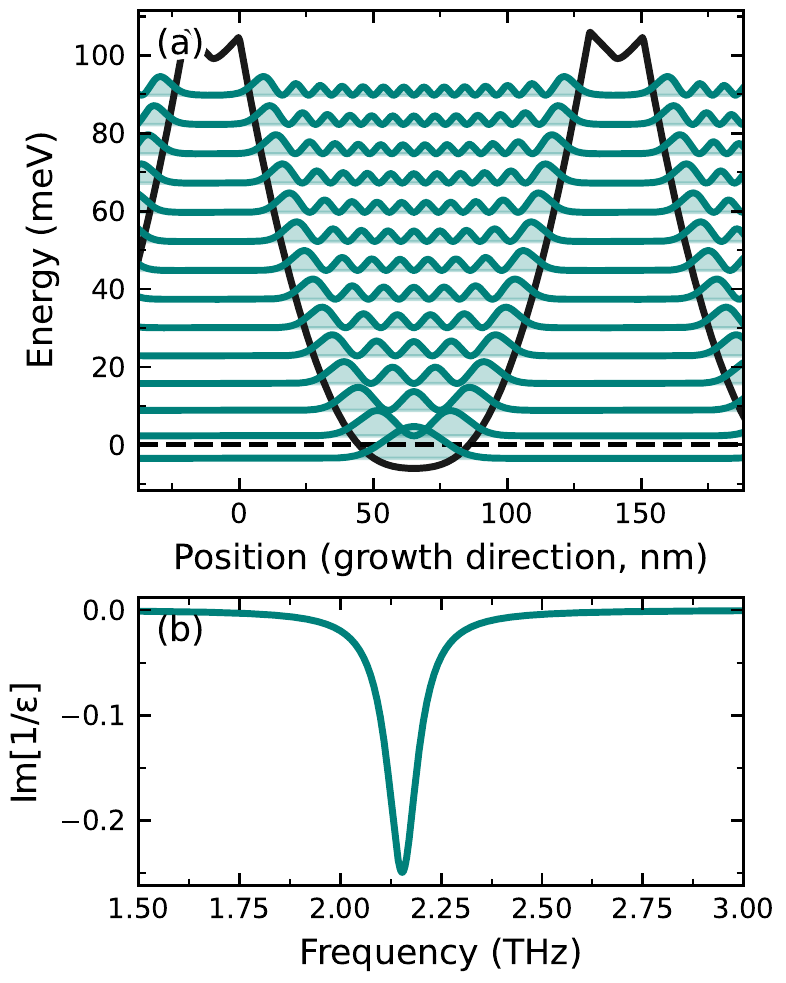}
  \caption{
    (a) Schr\"odinger-Poisson simulation of the parabolic quantum well with a continuously graded alloy.
    In black is the Conduction band energy, the green lines are the squared moduli of the wave functions \(\left| \Psi \right|^2\), while the dash line marks the energy position of the Fermi level.
    (b) Calculated absorption of the system with a linewidth of 10\%.
    }
  \label{fig:fig1}
\end{figure}

The heterostructure comprises a stack of 4 PQWs with 1.10\textsuperscript{11}\uspace cm\textsuperscript{-2} nominal sheet electron density per well.
The wells were realized in the Al\textsubscript{x}Ga\textsubscript{1-x}As material system by varying the aluminum composition x in the range 2\%-20\% along the growth direction.
The PQWs were each 130.75\uspace nm wide, separated by 20\uspace nm Al\textsubscript{0.2}Ga\textsubscript{0.8}As barriers where the Si doping is introduced as \(\delta\)-layers at the barrier center.
Two GaAs cap layers (50-nm thick) are placed on top and bottom of the structure.
The overall active region thickness is 783\uspace nm.
\textbf{Figure 1}a shows the Schr\"odinger-Poisson simulation of the grown structure performed with the commercial software nextnano\cite{birnerNextnanoGeneralPurpose2007} and the calculated absorption centered at 2.2\uspace THz (\textbf{Figure 1}b).

\subsection{Characterization of the optical performances}

We have first characterized the optical performance of the grown structure.
The sample is shaped in a multi-pass prism configuration with the facets polished at 45 degrees angle.
Ti/Au surface coating is made on the side of the epitaxial growth to force the boundary conditions at the semiconductor-metal interface and ensure a maximum of electric field on the parabolic quantum well repetition.
The waveguides were placed into a continuous-flow cryostat inside a Fourier transform infrared spectrometer (Bruker IFS66V), with polarized broadband THz light incident on the facet.
To reach a spectral resolution as low as 0.5\uspace cm\textsuperscript{-1}, the whole FTIR was placed under vacuum to remove the contribution of water absorption lines that are present even at low relative humidity levels \(\mathrm{(RH<1\%)}\).
Detection was performed with a liquid-He-cooled Si bolometer.
\textbf{Figure 2}a shows the color-map evolution of the transmittance as function of the temperature.
By modeling the ISB transition with a diagonal dielectric tensor as described in Refs.\cite{zaluznyCouplingInfraredRadiation1999,manceauMidinfraredIntersubbandPolaritons2014,knorrIntersubbandPolaritonpolaritonScattering2022}, the central frequency (\(\nu_0\)), the linewidth (\(\gamma_i\)) and the plasma frequency (\(\omega_p\)) can be extracted. Furthermore, one can gauge the evolution of the stack absorption as a function of the temperature. For a single well, the integrated 2D absorption is expressed as in \cite{helmBasicPhysicsIntersubband1999}:

\begin{equation}
  \int_{-\infty}^{+\infty}\alpha_{2D}(\omega)d\omega = \frac{\omega^{2}_{p}L_{QW}}{c\sqrt{\varepsilon_{s}}}
\end{equation}

Hence, one can define the parameter \(\alpha = \sqrt{N_{QW}}\omega_p\), where \(N_{QW}\) is the number of wells, and monitor the stack absorption strength as a function of the temperature.

The evolution of the linewidth (\(\gamma_i\)) and the central frequency (\(\nu_0\)) are plotted in Figure 2b and c respectively. We first note a very stable single absorption peak at a frequency around 2.1\uspace THz over the whole temperature range.
The linewidth shows stronger temperature dependence, with a record value of 69\uspace GHz at 8\uspace K (3.2\% FWHM) that remains below 10\% up to 150\uspace K.
Despite having several subbands populated at higher temperature, the parameter alpha (Figure 2d) is stable over the whole temperature range, evidence that the number of electrons participating in the absorption remains mostly constant thanks to the harmonicity of the system.

\begin{figure}
  \includegraphics{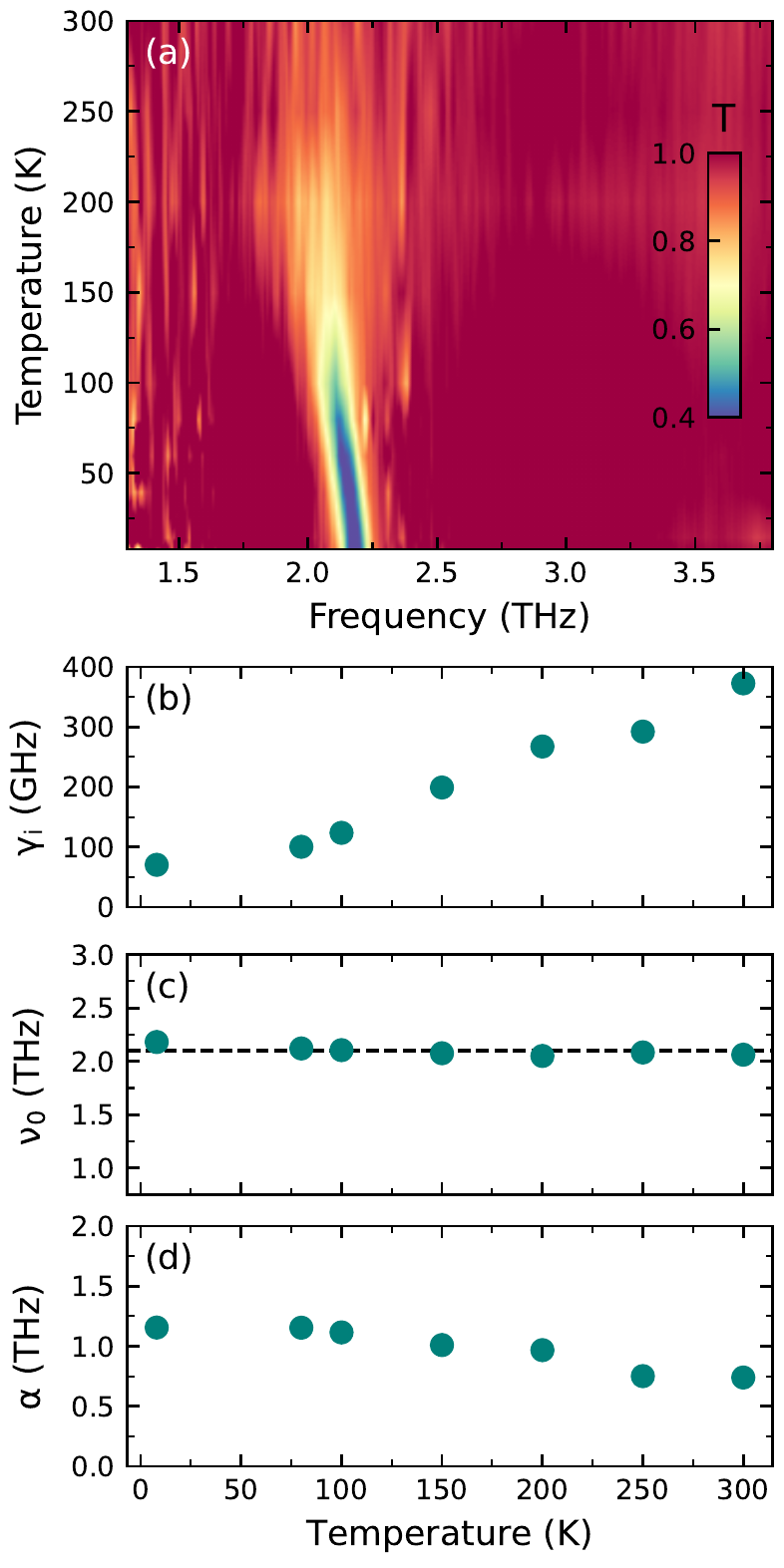}
  \caption{
    (a) Transmittance measurement of the grown heterostructure as function of the temperature.
    (b) Extracted linewidth using Voigt fitting formula as function of the temperature. We measured 69\uspace GHz linewidth at 10\uspace K.
    (c) Central frequency of the transition as function of the temperature.
    The dashed line is a guide for the eye to show the stability of the transition central frequency (d) Integrated absorption of the transition as function of the temperature.
    }
  \label{fig:fig2}
\end{figure}

\section{Ultra-Strong light-matter coupling in microcavity}

\begin{figure*}
  \includegraphics{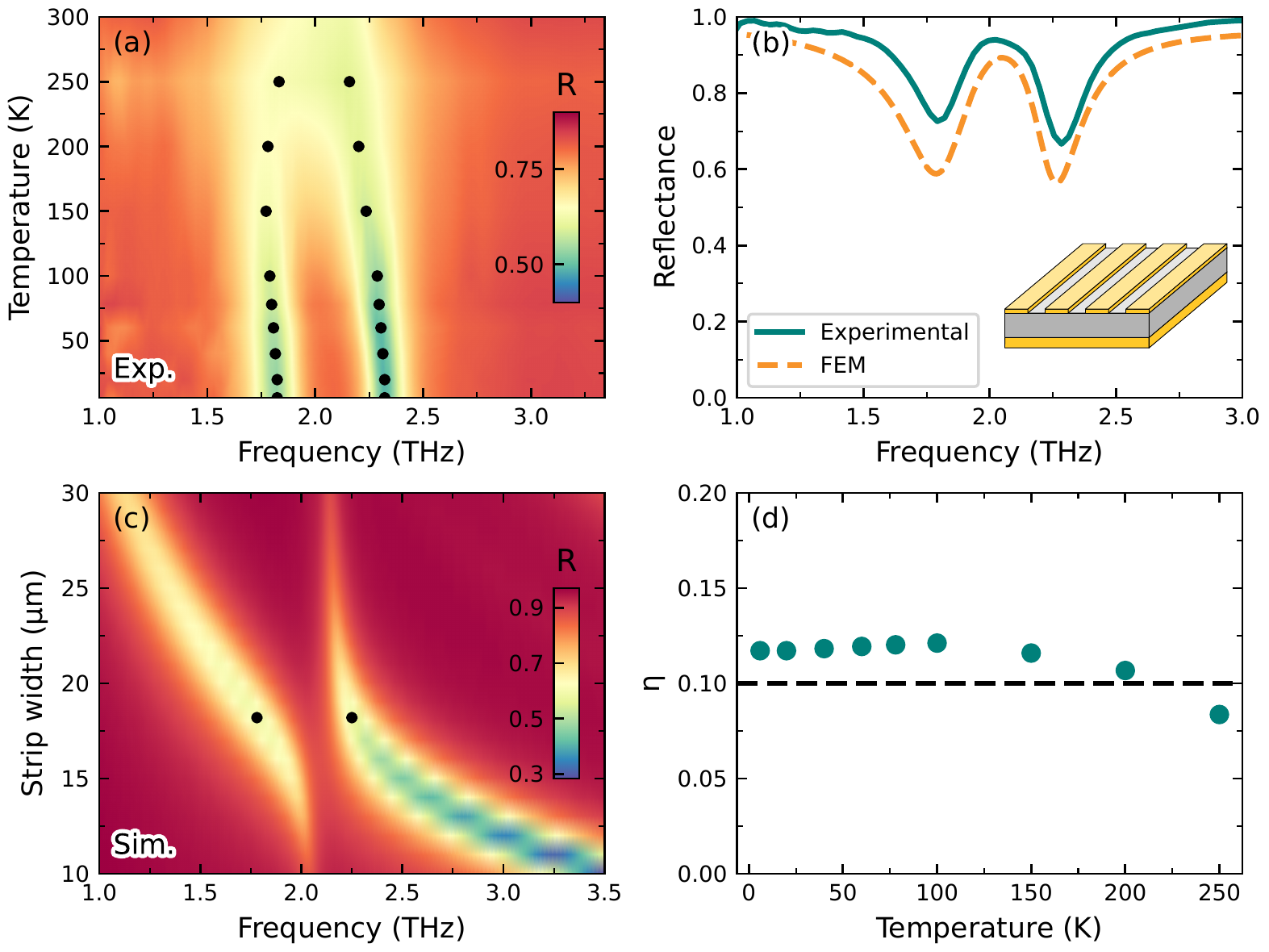}
  \caption{
    (a) Experimental reflectance of the ISB polaritonic system as function of the temperature.
    The dots mark the positions of the reflectance minima (b) Fit (dashed line) of the experimental reflectance at 78\uspace K (full line), using finite element simulations. In inset is a schematic of the micro-ribbon resonator.(c) Simulation of the polaritonic system dispersion, attesting that the system is at the anti-crossing.
    The dots mark the experimental positions of the polaritonic peaks in the sample having a ribbon size of 18\uspace \textmu m.
    (d) Coupling strength parameter (\(\eta = \Omega_R/\omega_0\)) as function of the temperature. The dashed line marks the limit of the ultra-strong coupling regime.
    }
  \label{fig:fig3}
\end{figure*}

We then explore operation in the strong light-matter coupling regime.
To this scope, the active region is placed within non-dispersive metal-insulator-metal (MIM) ribbon micro-resonators.\cite{todorovOpticalPropertiesMetaldielectricmetal2010}
The details of the micro-fabrication process can be found elsewhere.\cite{manceauOpticalCriticalCoupling2013}
We have fabricated several samples with different sizes of metallic ribbon that we have measured in reflectance at 300\uspace K.
For each samples, the minima of reflectance from the different resonant photonic modes (\(\mathrm{TM_{01}, TM_{03}}\)) were placed on the dispersion plot obtained from a full EM simulation along with the experimentally measured position of the ISB transition. 
Hence, we could define the sample placed at the anti-crossing that allows measuring the Rabi splitting and accurately estimate the Rabi frequency.
The lateral size of the metal stripe, \(\Lambda\), is 18\uspace \textmu m, which sets the resonator operating frequency.
The gap in between each stripe, D, is equal to 4\uspace \textmu m. 
\textbf{Figure 3}a shows the color-map evolution of the reflectance as function of the temperature.
One can clearly see the lift of degeneracy between cavity and ISB transition that occurs already below 250\uspace K, and the two polaritonic branches gaining in contrast while the temperature decreases.
To infer the Rabi frequency of the strongly coupled system, we have performed careful EM simulations to fit the measured Rabi splitting, where the ISB transition is depicted using the diagonal dielectric tensor approach aforementioned.
In the quantum well plane, the material is described by the GaAs permittivity with phonons and a Drude model accounting for the free motion of the electrons.
Doing so, the only fitting parameter is the sheet doping density (\(\mathrm{n_{2D}}\)) that enters the plasma frequency expression:

\begin{equation}
  \omega_p = \sqrt{\frac{f_{12} e^2 n_{2D}}{\varepsilon_0 \varepsilon_r m^* L_{QW}}}
\end{equation}

Here \(f_{12}\) is the oscillator strength (equal to 0.97 for a PQW) between the first and second subband, \(m^*\) is the GaAs effective mass and \(L_{QW}\) is the length of the period comprising one well and one barrier.
Figure 3b shows the experimental fitted reflectance at 10\uspace K, where the \(n_{2D}\) is equal to 4.10\textsuperscript{10}\uspace cm\textsuperscript{-2}, a factor two below the nominal doping.
This discrepancy with the nominal doping value might be due to the depletion of the two QWs close to the metal-semiconductor interface where strong band bending occurs.
Based on the fitted permittivity, we ran a full simulation with different metal stripe sizes in order to reconstruct the dispersion and confirm that our microcavity is indeed at the anti-crossing condition (Figure 3c).
It is important to note that experimentally, we have access to the Rabi splitting frequency, expressed as \(\Omega_{\mathrm{splitting}} = 2\Omega_R \sqrt{\Gamma_{\mathrm{opt}}}\).
Here, \(\mathrm{\Gamma_{opt}}\) is a dimensionless parameter, which represents the fraction of electromagnetic energy coupled to the out-of-growth-plane component of the electric field and spatially overlapping the active region, as described in Ref.\cite{zanottoAnalysisLineShapes2012}.
In turn, isolating the Rabi frequency in the above expression, and using \(\mathrm{\Gamma_{opt}}\)=0.95 (obtained from the finite element method simulations), we estimate that \(\Omega_R\) = 0.25\uspace THz.
We can extract \(\Omega_R\) along with \(\omega_0\) for the whole set of temperatures and estimate the dimensionless parameter that gauges the coupling strength \(\eta = \Omega_R / \omega_0\).
Figure 3d demonstrates that the system operates in the ultra-strong coupling regime up to 200\uspace K. Furthermore, we have extracted the linewidth of the polaritonic states,  showing $\gamma_{pol}$ of around 180\uspace GHz and 140\uspace GHz up to 78K (for lower and upper polaritonic states respectively). This is equivalent to Q factors of 10 and 17, which is about 5 times higher than the one of Landau polaritons. Such performance is particularly important in view of non-adiabatic switching experiments since the emission rate of vacuum quanta can be significantly increased by a narrower polariton linewidth, as thoroughly discussed in. \cite{deliberatoQuantumVacuumRadiation2007} We also anticipate that the increase in the transition coherence time shall improve the signal to noise ratio within the coherent electro-optic sampling technique, in comparison to the incoherent background radiation inherently present.\cite{moskalenkoParaxialTheoryDirect2015,riekDirectSamplingElectricfield2015,benea-chelmusSubcycleMeasurementIntensity2016,benea-chelmusElectricFieldCorrelation2019} Another way to gauge the importance of improved polaritons linewidth is to use the figure of merit defined in \cite{RevModPhys.91.025005}, that is \(U = \sqrt{\eta C}\), where $\eta$ is the coupling strength and C the cooperativity defined as \(C = 4\Omega_R^2/\kappa\gamma_{i}\), where $\kappa$ and  $\gamma_{i}$ are the cavity and transition linewidth respectively. A value of U$\geq$1 indicates that the system operates in the USC regime with a high degree of coherence, allowing to access its physics without being hindered by decoherence. For the present work, even though the coupling strength is relatively small, we estimate U equals to 1.7, thanks to the improved transition linewidth. In our future work, we will aim at improving the coupling strength by lowering the transition to 1.1 THz while increasing the doping density by a reasonable factor 3 ($\sqrt{3}$ on $\Omega_R$). This would lead to an increase of $\eta$ by a factor 3.5 and in turn to U = 6.7 if the ISB transition linewidth is preserved.  Such value would in fact be equivalent to the one reported in the literature for Landau polaritons, attesting that our material system is an alternative to it.

\section{Ultra-Strong light matter coupling in sub-wavelength resonators}

\begin{figure*}
  \includegraphics{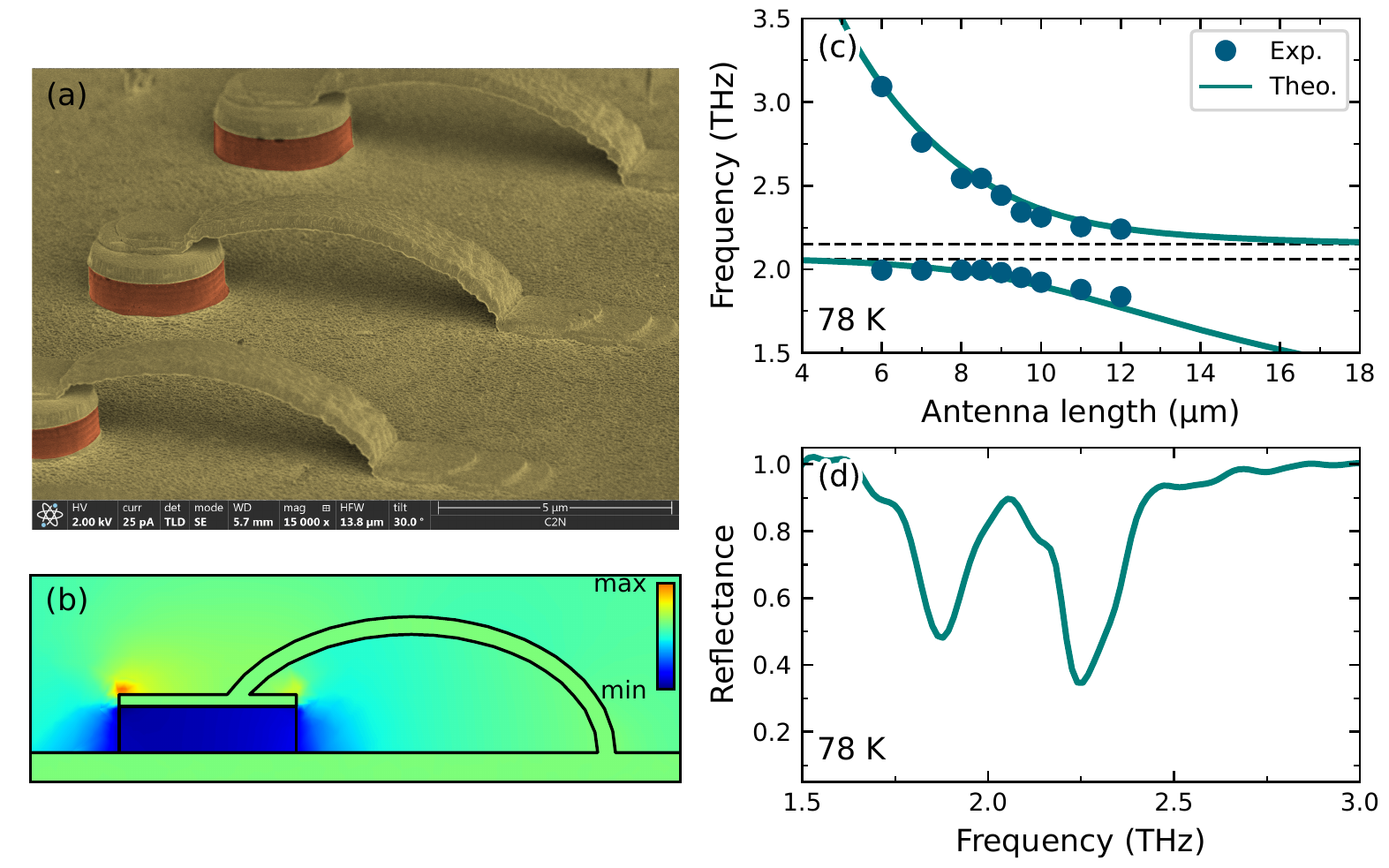}
  \caption{
    (a) Scanning electron microscope image of the LC resonators array.
    Colorized in red is the capacitive section hosting the parabolic quantum wells, the rest being gold (b) Simulated confined electric field in the direction orthogonal to the growth plane (Ez).
    (c) Dispersion of the polaritonic system as a function of the antenna length. 
    Dashed lines are the secular equation while the dots are the experimental points.
    The dotted parallel lines mark the edges of the polaritonic gap.
    (d) Experimental reflectance of the array with antenna length of 11\uspace \textmu m at a temperature of 78\uspace K, attesting of the excellent optical performances of the system.
  }
  \label{fig:fig4}
\end{figure*}

Finally, we have verified the compatibility of this active region with recently developed deeply sub-wavelength microcavities. 
The idea is to use a cavity that operates in the high-frequency inductor-capacitor (LC) resonant mode and which confines the EM field within a sub-wavelength volume.\cite{paulilloCircuittunableSubwavelengthTHz2014,waltherMicrocavityLaserOscillating2010}
Such resonators have been successfully used in recent years to demonstrate strong-light matter coupling with a reduced number of electrons \cite{jeanninUltrastrongLightMatter2019} and to increase the operating temperature of QWIPs.\cite{jeanninHighTemperatureMetamaterial2020}
More recently, in pursuit of non-adiabatic cavity modulation, the implementation of a frequency switching functionality of 280 GHz in less than 200 fs was demonstrated by taking advantage of the circuital properties of the device.\cite{goulainFemtosecondBroadbandFrequency2021}
\textbf{Figure 4}a shows a scanning electron microscope (SEM) picture of the fabricated device.
The fabrication details can be found in Ref.\cite{paulilloCircuittunableSubwavelengthTHz2014}.
The capacitor section (colorized red) is a cylinder of 1.5\uspace \textmu m diameter by 783\uspace nm height, with an effective volume of \(2.10^{-6} \lambda_0^3\).
Since the resonance frequency scales with \(\sqrt{LC}\), we have fabricated several samples with different antenna lengths to reconstruct the polaritonic dispersion (Figure 4b). 
From the reflectivity minima, we can identify the two polariton branches and fit them with the secular equation\cite{todorovUltrastrongLightMatterCoupling2010}

\begin{equation}
  \left( \omega - \omega_c^2 \right) \left( \omega - \tilde{\omega}_{21}^2 \right) = \Gamma_{\mathrm{opt}}\omega_p^2 \omega_c^2
\end{equation}

Where \(\omega_c\) is the cavity resonance numerically calculated, \(\tilde{\omega}_{21}^2\) is the measured transition while \(\Gamma_{\mathrm{opt}}\) and \(\omega_\mathrm{p}\) were defined above.
The dotted lines are placed along the asymptotes and reveal the presence of a polaritonic gap, a key signature of the USC regime.
Figure 4c is the reflectance measurement at the anti-crossing point where we measure a Rabi splitting \(\Omega_s\) = 0.37\uspace THz.
The optical confinement within the LC resonator is lower than the MIM cavities, \(\Gamma_{\mathrm{opt}}\) = 0.83 in this case.
We estimate a Rabi frequency of 0.24\uspace THz and a coupling strength \(\eta\) of 0.11, confirming that the system operates in the USC regime.
Since the electron sheet density is \(\mathrm{n_{2D} = 4.10^{10}}\)\uspace cm\textsuperscript{-2}, we estimate that the USC is achieved with around 3000 electrons per resonator coupled to the quantized EM mode.
Even lower numbers could be reached by reducing the size of the device capacitance.
For instance, considering a single PQW active region and leveraging the higher resolution offered by the electron beam lithography, a resonator with a capacitive section of 500\uspace nm diameter over 315\uspace nm height could be fabricated.
With such a device, we estimate that the USC could be achieved with only 400 electrons per resonator, a further step toward the study of the USC with few electrons.\cite{todorovFewElectronUltrastrongLightMatter2014,kellerFewElectronUltrastrongLightMatter2017}

\section{Conclusion}

In conclusion, we have used continuously graded parabolic QWs with outstanding optical performances to demonstrate ISB polaritons around 2\uspace THz.
The intrinsic low frequency bottleneck of SQWs is overcome thanks to the robustness of the harmonic oscillator allowing the formation of a polariton mode as low as 1.8\uspace THz with Q factors up to 17.
The system operates in the ultra-strong light-matter coupling regime up to 200\uspace K.
We further demonstrate that the USC regime is maintained in deeply sub-wavelength volumes where around 3000 electrons per resonator are involved in the interaction.
A further reduction towards hundreds of electrons could be envisioned using state-of-the-art nanofabrication processing.
Finally, further optimization of the system cooperativity can be envisioned along with the implementation of an ultrafast switching element as in Ref.\cite{goulainFemtosecondBroadbandFrequency2021} that would make such system a more practical alternative to Landau polaritons in the study of quantum vacuum radiation at long-wavelengths.

\medskip
\textbf{Acknowledgements} \par 

This work was supported by the European Union Future and Emerging Technologies (FET) Grant No. 737017 (MIR-BOSE), and from the French National Research Agency, project TERASEL (ANR-18-CE24-0013).
This work was partially supported by the French RENATECH network, Canada First Research Excellence Fund, and the Natural Sciences and Engineering Research Council of Canada (NSERC).
We thank C. Ciuti for granting access to Paris Cit\'e University cleanroom, and P. Filloux and S. Suffit for help.

\medskip

%

\bibliographystyle{MSP}
\bibliography{biblio}

\end{document}